\documentclass[letterpaper]{article}
\usepackage{aaai}
\usepackage{times}
\usepackage{helvet}
\usepackage{courier}

\begin{document}
\nocopyright

\title{
Naming Games in Spatially-Embedded Random Networks
\thanks{We thank B. Yener, J.W. Branch, and A. Barrat for comments on this
work. This research was supported in part by NSF Grant Nos.\ DMR-0426488 (G.K. and Q.L.),
NGS-0103708 (B.K.S. and Q.L.), and by Rensselaer's Seed Program.
B.K.S. was also supported through participation in the International Technology Alliance sponsored
by the U.S. Army Research Laboratory and the U.K. Ministry of Defence.}
}

\author{Qiming Lu \and G. Korniss \\
Department of Physics\\
Rensselaer Polytechnic Institute\\
Troy, New York 12180--3590\\
luq2@rpi.edu, korniss@rpi.edu
\And
Boleslaw K. Szymanski \\
Department of Computer Science\\
Rensselaer Polytechnic Institute\\
Troy, New York 12180--3590\\
szymab@rpi.edu}

\maketitle

\begin{abstract}
We investigate a prototypical agent-based model, the Naming Game,
on random geometric networks. The Naming Game is a minimal model,
employing local communications that captures the emergence of
shared communication schemes (languages) in a population of
autonomous semiotic agents. Implementing the Naming Games on
random geometric graphs, local communications being local
broadcasts, serves as a model for agreement dynamics in
large-scale, autonomously operating wireless sensor networks.
Further, it captures essential features of the scaling properties
of the agreement process for spatially-embedded autonomous agents.
We also present results for the case when a small density of
long-range communication links are added on top of the random
geometric graph, resulting in a ``small-world"-like network and
yielding a significantly reduced time to reach global agreement.
\end{abstract}

\section{Introduction}

Reaching agreement without global coordination is of fundamental
interest in large-scale autonomous multi-agent systems. In the
context of social systems, the objective is to understand and
predict the emergence of large-scale population-level patterns
arising from empirically supported local interaction rules between
individuals (e.g., humans). Examples for such phenomena driven by
social dynamics include the emergence and the evolution of
languages \cite{Nowak_JTB1999,Nowak_Science2001,Nowak_PNAS2004} or
opinion formation
\cite{Castellano2005,Eli,Deffuant,Durlauf,KR2003,Sznajd}. From a
system-design viewpoint in technological (e.g., sensor) networks
\cite{Lee_2005,Collier_2004}, the objective can be somewhat
reversed, in that it is to construct local rules giving rise to a
fast and efficient convergence to a global consensus, when needed.

In this paper we consider and slightly modify a simple set of
rules, referred to as Language or Naming Games (NG), originally
proposed in the context of semiotic dynamics \cite{Steels,Kirby}.
Such problems have become of technological interest to study how
artificial agents or robots can invent common classification or
tagging schemes from scratch without human intervention
\cite{Steels,Kirby}. The original model
\cite{Steels,Steels_1998,Steels_1995,Steels_1997} was constructed
to account for the emergence of shared vocabularies or conventions
in a community of interacting agents. More recently, a simplified
version of the NG was proposed and studied on various network
topologies by Baronchelli et al.
\shortcite{Baronchelli_2005a,Baronchelli_2005b,Baronchelli_2006a},
and by Dall'Asta et al. \shortcite{Baronchelli_2006b} The
advantage of studying a minimal model is that one can gain a
deeper understanding of the spontaneous self-organization process
of networked autonomous agents in the context of reaching global
agreement, and can extract quantitative scaling properties for
systems with a large number of agents. This simplified version of
the NG was investigated on fully-connected (FC) (also referred to
as mean-field or homogeneous mixing)
\cite{Baronchelli_2005a,Baronchelli_2005b}, regular
\cite{Baronchelli_2006a}, and small-world (SW) networks
\cite{Baronchelli_2006b}.

In the FC network, each agent has a chance to meet with all others
and compare their current local vocabularies before updating them.
On regular networks, agents have only a limited and fixed number
of neighbors with whom they can interact/communicate. The
communication in both cases is ``local", in that {\em pairs of
agents} are selected to interact and to update their vocabularies.
The basic algorithmic rules of the NG are as follows
\cite{Baronchelli_2005a,Baronchelli_2006a}. A pair of
``neighboring" nodes (facilitated by the underlying communication
topology), a ``speaker" and a ``listener", are chosen at random.
The speaker will transmit a word from her list of ``synonyms" to
the listener. If the listener has this word, the communication is
a success, and both players delete all other words, i.e., collapse
their list of synonyms to this one word. If the listener does not
have the word transmitted by the speaker, she adds it to her list
of synonyms without any deletion. It was found that employing the
above local rules ({\em pair-wise} interactions), after some time,
the agents vocabularies converge to a unique vocabulary shared
among all agents
\cite{Baronchelli_2005a,Baronchelli_2005b,Baronchelli_2006a,Baronchelli_2006b}.
The major differences between the NG on FC and on regular (e.g.,
two-dimensional) networks arise in the memory needed to develop
the common language before convergence occurs, and in time $t_c$ needed to
reach global agreement.
In the FC network, the convergence
process to global agreement is fast [$t_c\sim{\cal O}(N^{1/2})$
for $N$ agents], but large memory is needed per agent
\cite{Baronchelli_2005a}. For a regular two-dimensional network
(or grid), spontaneous evolution toward a shared dictionary is
slow [$t_c\sim{\cal O}(N)$], but the memory requirement is much
less severe \cite{Baronchelli_2006a}. When the NG is implemented
on Watts-Strogatz \shortcite{Watts98} SW networks, the agreement
dynamics performs optimally in the sense that the memory needed is
small, while the convergence is much faster than on the regular
networks [$t_c\sim{\cal O}(N^{0.4})$, close to that of the FC
network] \cite{Baronchelli_2006b}.

The situation described above (i.e., the need for shared
vocabularies) can also be quite realistic in the context of sensor
networks \cite{Lee_2005,Collier_2004}. Envision a scenario where
mobile or static sensor nodes are deployed in a large
spatially-extended region and the environment is unknown, possibly
hostile, the tasks are unforeseeable, and the sensor nodes have no
prior classification scheme/language to communicate regarding
detecting and sensing objects. Since subsequent efficient
operation of the sensor network inherently relies on unique object
identification, the autonomous development of a common ``language"
for all nodes is crucial at the exploration stage after network
deployment \cite{Collier_2004,Lee_2005}. For this task, however,
there are more efficient and faster schemes, guaranteeing to reach
global agreement on the naming (tagging) of an object. In
particular, basic leader-election (LE) algorithms
\cite{Angluin,HS,LeLann,MWV,VDIKT} could be employed to arrive
at a common word among a community of agents which observed the
object to be named: Upon observation, each agent ``coins" a random
tag (identification number) for the object. Following the
observation, the observing agents participate in the leader
election algorithm (with not the purpose of electing a leader, but
choosing a unique identifier for the object).  For example, in a
two-dimensional regular network of $N$ agents, the convergence to
a unique identifier takes time of ${\cal O}(\sqrt{N})$ on average.
Convergence time in the NG in two dimensions is of ${\cal O}(N)$,
significantly longer than the LE algorithm when $N$ is large.
Thus, for the purpose of constructing a shared classification or
tagging scheme in a sensor network, launching the LE algorithm is
the preferable choice. Unlike social networks, sensor networks,
although operating autonomously, are intelligently designed (by
humans), who can make the choice a priori which algorithm to
employ based on their efficiency.

There are possible situations, however, when the NG algorithm, in
addition to being interesting for its own merit in studying
agreement dynamics on various networks, can also be beneficial
from a system-design viewpoint. That can be the case when one does
not intend the outcome of the agreement to be easily predictable.
The actual process of electing a ``leader" or coordinator among
sensor nodes may actually be such a scenario. The leader must
typically be a trusted node, with possible responsibilities
ranging from routing coordination to key distribution
\cite{DeCleene}. The basic LE algorithms are essentially based on
finding global extremum (e.g., maximum) through local
communications \cite{Angluin,HS,LeLann}. Thus, the elections can
be stolen by placing a node in the network with a sufficiently
high ID (e.g., the largest number allowed by the number
representation scheme of the sensor chips.) Along these lines, a
possible application of the NG algorithm is autonomous key
creation or selection for encrypted communication in a community
of sensor nodes. Instead of having a centralized or hierarchial
key management system with domain and area key distributors
\cite{DeCleene}, group of sensor nodes can generate a shared
``public" key (becoming visible to group members only).

Sensor networks are both spatial and random. As a large number of
sensor nodes are deployed, e.g.,  from vehicles or aircrafts, they
are essentially scattered randomly across large spatially-extended
regions. In the corresponding abstract graph, two nodes are
connected if they mutually fall within each others transmission
range, depending on the emitting power, the attenuation function
and the required minimum signal to noise ratio. Random geometric
graphs (RGGs), also referred to as spatial Poisson/Boolean graphs,
capturing the above scenario, are a common and well established
starting point to study the structural properties of sensor
network, directly related to coverage, connectivity, and
interference. Further, most structural properties of these
networks are discussed in the literature in the context of
continuum percolation \cite{percolation,penrose,Dall_2002}.

The common design challenge of these networks is to find the
optimal connectivity for the nodes: If the connectivity of the
nodes is too low, the coverage is poor and sporadic. If the node
connectivity is too high, interference effects will dominate and
result in degraded signal reception
\cite{Kumar2000,Kumar2004,phtr_networks,krause04,BCS}. From a
topological viewpoint, these networks are, hence, designed to
``live" somewhere above the percolation threshold. This can be
achieved by adjusting the density of sensor nodes and controlling
the emitting power of the nodes; various power-control schemes
have been studied along these lines \cite{Kumar2000,krause04,BCS}.
In this paper we consider RGGs above the percolation threshold, as
minimal models for the underlying network communication topology.
Further, we consider RGGs with an added small density of ``random"
long-range links. The resulting structure resembles small-world
(SW) networks \cite{Watts98,NEWMAN_SIAM}, also well studied in the
context of artificial \cite{KNGTR03a,LKS_INSS2006} and social
systems \cite{Watts99,NEWMAN_SIAM}.The focus of this work is to
study the NG algorithm {\em on} these well studied graphs.


\section{Naming Games on Random Geometric Networks}

As mentioned above in the Introduction, first we consider random
geometric graphs in two dimensions
\cite{percolation,penrose,Dall_2002} as the simplest topological
structures capturing the essential features of ad hoc sensor
networks. $N$ nodes are uniformly random distributed in an
$L$$\times$$L$ spatial area. For simplicity we consider identical
radio range $R$ for all nodes. Two nodes are connected if they
fall within each other's range. An important parameter in the
resulting random geometric graph is the average degree
$\overline{k}$ (defined as the average number of neighbors per
node), $\overline{k}$$=$$2K/N$, where $K$ is the total number of
links and $N$ is the number of nodes. In random geometrical
networks, there is a critical value of the average degree,
$\overline{k}_c$, above which the largest connected component of
the network becomes proportional to the total number of nodes (the
emergence of the giant component)
\cite{percolation,penrose,Dall_2002}. For two-dimensional RGGs
$\overline{k}_c$$\approx$$4.5$ \cite{Dall_2002}. There is a simple
relationship between the average degree $\overline{k}$, the
density of nodes $\rho$$=$$N/L^2$, and the radio range $R$ of the
nodes \cite{percolation,penrose,Dall_2002},
$\overline{k} =\rho \pi R^2$,
which can be used to control the connectivity of the network.

\subsection{The Naming Game}

We consider the Naming Game on random geometrical graphs. In the
original context of the NG, agents try to reach agreement in
finding a unique ``word" for an object observed by them. In one of
the above proposed potential applications, agents try to generate a shared
public key for encrypted communication. For simplicity, we will
use the term ``word" for the latter as well when describing the
algorithm.

We modify the communication rules to make them applicable for
sensor networks. Instead of pairwise communications, nodes will
initiate {\em broadcast} (to all neighbors) in a continuous-time
asynchronous fashion. In this paper we consider the initial
condition when the ``vocabulary" of each node is empty. At every
elementary time step, a node is chosen randomly out of $N$ nodes
(mimicking Poisson asynchrony for large $N$). This node (the
``speaker") will broadcast a word from her list of ``synonyms"; if
her list of synonyms is empty, the speaker randomly invents a
word; if she already has several synonyms, it randomly chooses
one. Her neighbors (the ``listeners") compare their vocabularies
with the word transmitted by the speaker. If a listener has this
word, she considers the communication a success, and she deletes
all other words, collapsing her list of synonyms to this one word.
If a listener does not have the word transmitted by the speaker,
she adds it to her list of synonyms without any deletion. If at
least one listener had the word transmitted, the speaker considers
it (at least a partial) success, and (somewhat optimistically)
collapses her list of synonyms to this one word. At every step,
the ``success" rate $S$ is defined as the fraction of listeners
who were successful (i.e., those that had the word transmitted by
the speaker). From the above it is clear that the listeners have
to report the outcome of the ``word matching" to the speaker,
hence the elementary algorithmic step requires $(\overline{k}+1)$
broadcasts. In this paper time $t$ is given in units of
$(\overline{k}+1)$ broadcasts per node (during which, on average,
$N$ word matching have been attempted). The main difference
between the above algorithm and the one in the works by
Baronchelli et al. is the {\em broadcast} (instead of pairwise
communications) and the underlying network (RGG in this paper) to
capture the essential features of the NG in sensor networks.

Other initial conditions, e.g., applicable to public key
generation for a community of networked agents, can also be
employed; instead of starting from ``scratch" (empty list of words
for each node), each agent has a pre-generated (possibly long)
list of words. The different initial conditions only have an
effect on the early-time behavior of the system. For this
scenario, results will be presented elsewhere.

\begin{figure}[t]
\vspace*{2.0truecm}
       \includegraphics{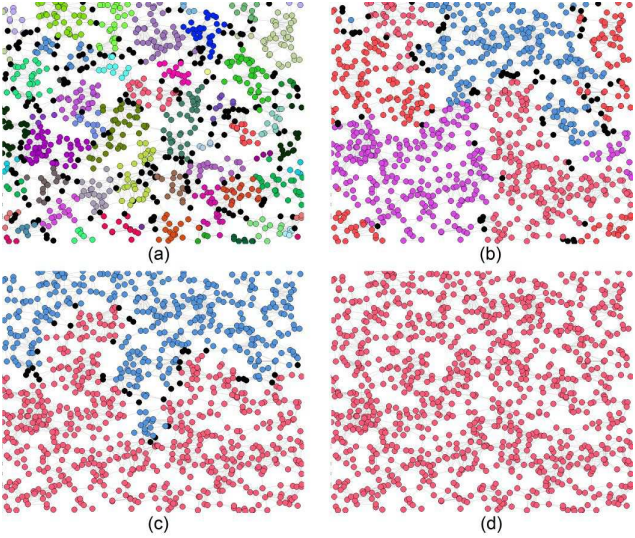}
\vspace*{5.5truecm}
\caption{(Color figure) Snapshots of the time evolution of the contents
of the agents' word lists during the process of reaching global agreement
on RGG for $N=1,000$ nodes at time
(a) $t=1$;
(b) $t=43$;
(c) $t=169$;
(d) $t=291$.
The average degree is $\overline{k}$$\approx$$12$. Initially, the word lists are
empty for all agents. Time is measured in units of
$(\overline{k}+1)$ broadcasts per node. Different colors correspond
to different words, with black indicating nodes with multiple words.
After the early-time increase in the number of different words in
the systems, small spatial clusters sharing the same word quickly
form, then subsequently ``coarsen" until eventually only one global
cluster prevails.}
\label{fig.snapshots}
\end{figure}

\begin{figure}[tb]
\vspace*{2.0truecm}
       \includegraphics{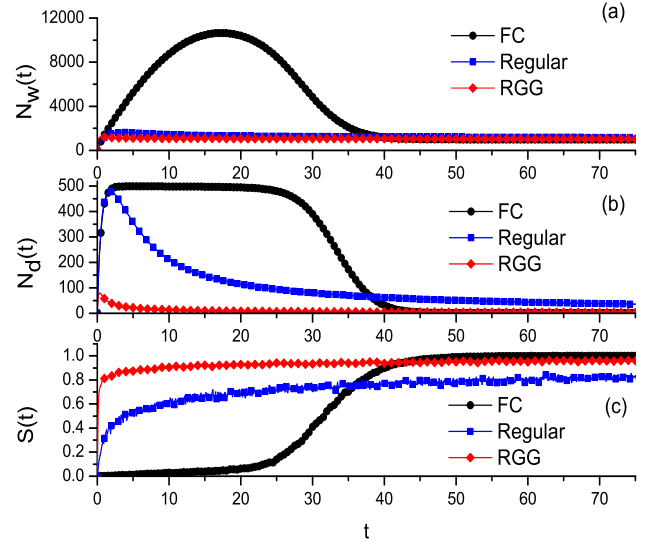}
\vspace*{6.0truecm}
\caption{Time evolution of the relevant
observables in the Naming Game in the fully-connected (FC),
two-dimensional regular (with four nearest neighbors), and random geometric networks
(RGG) for $N$$=$$1024$, averaged over $1,000$ independent network
realizations; (a) the total number of words in the
system $N_{w}(t)$; (b) the number of different words $N_{d}(t)$;
(c) the average success rate $S(t)$. The average degree of the
underlying RGG is $\overline{k}$$\approx$$12$. Data for the FC and
$2d$ regular networks are reproduced by our simulations, following
Refs.~\cite{Baronchelli_2005a,Baronchelli_2006a}, for comparison.}
\label{fig.mf-rgg-reg}
\end{figure}

When starting from empty vocabularies, agents invent words
randomly. After time of ${\cal O}(1)$ [on average of order
$(\overline{k}+1)$  broadcast per node], ${\cal
O}(N/(\overline{k}+1))$ different words have been created.
Following the early-time increase of the number of different words
$N_d(t)$, through local broadcasts, agents slowly reconcile their
``differences", and eventually will all share the same word.
First, a large number of small spatial clusters sharing the same
word develop. By virtue of the random {\em diffusive} motion of
the interfaces separating the clusters, more and more of the small
clusters are being eliminated, giving rise to the emergence of
larger clusters, eventually leading to one cluster in which all
nodes are sharing the same word. As suggested by Baronchelli et
al. \shortcite{Baronchelli_2006a}, this late-time process is
analogous to coarsening, a well-known phenomenon from the theory
of domain and phase ordering in physical and chemical systems
\cite{Bray}. Figure~\ref{fig.snapshots} shows snapshots of
vocabularies of the nodes at different times. For later times,
group of nodes which already share the same word, slowly coarsen,
until eventually only one domain prevails. This behavior is also
captured by Fig.~\ref{fig.mf-rgg-reg}(b), tracing the number of
different words as a function of time $N_d(t)$, eventually
reaching global agreement, $N_d=1$.

\subsection{Basic Scaling Considerations and the Analogy with Coarsening}
Before turning to the detailed discussion of our simulation
results, we first sketch the framework of coarsening theory
\cite{Bray}, applicable to the observed late-time dynamics of the
NG on regular $d$-dimensional lattices \cite{Baronchelli_2006a}.
Coarsening has also been observed in other models relevant to
opinion formation and social dynamics \cite{BNFK1996,KR2003}.
Unlike other minimalist (typically two-state) models often
employed to study opinion formation \cite{Durlauf}, such as the
one studied by Sznajd-Weron \& Sznajd \shortcite{Sznajd}, the
Voter model \cite{intpart,BNFK1996}, or the majority rule model
\cite{KR2003}, in the NG, each agent can be in an {\em unlimited}
number of discrete states (corresponding to the chosen word).
Further, at any instant before reaching global consensus, an agent
can have different possible words for the object. Because of the
potentially unlimited number of discrete states the agents can
``reside in", we believe
 [this belief is also supported by preliminary data \cite{Wykes}] that the full
coarsening characteristics exhibited by the NG will match that of
the infinite-state Potts model
\cite{Majumdar1995a,Majumdar1995b,bAvraham2000}. After the initial
fast local ordering, essentially each cluster corresponds to a
{\em different} state (word). Hence, through the coarsening
process, domain walls (interfaces) separating different clusters
will {\em coalesce} (as opposed to domain-wall {\em annihilation}
in two-state models).

While RGG is a random structure, it is embedded in two dimensions,
and we also attempt to employ elementary scaling arguments from
coarsening theory. According to Ref.~\cite{Baronchelli_2006a}, on
regular $d$-dimensional lattices, the typical size of domains
(each with already agreed upon one word) is governed by a single
length scale $\xi(t)\sim t^{\gamma}$ with $\gamma$$=$$1/2$,
analogous to that of domain formation in systems with a
non-conserved order parameter \cite{Bray}. Thus, in $d$
dimensions, the total number of {\em different} words $N_d$ at
time $t$ scales as the typical number of domains
\begin{equation}
N_d(t) \sim \frac{N}{\xi^d(t)}\sim \frac{N}{t^{d\gamma}} \;.
\label{Nd}
\end{equation}
Further, the total number of words $N_w$ ($N_w/N$ being the average memory load per agent),
at this late coarsening
stage, can be written as the number of nodes $N$ plus the number
of nodes with more than one (on average, between one and two) words,
separating the different domains. It is of order of typical number of domains
times the typical length of the interface of one domain, yielding
\begin{equation}
N_w(t)- N \sim \frac{N}{\xi^d(t)}\xi^{d-1}(t) \sim
\frac{N}{\xi(t)} \sim \frac{N}{t^{\gamma}}\;.
\label{Nw}
\end{equation}
Similarly, the ``failure rate" for word matching, $1$$-$$S(t)$,
(where $S(t)$ is the success rate) scales as the fraction of nodes
at the interfaces separating domains with different words
\begin{equation}
1-S(t) \sim \frac{1}{\xi(t)} \sim \frac{1}{t^{\gamma}}\;.
\label{S}
\end{equation}
The main feature of the above power-law decays (up to some
system-size dependent cut-offs) is that the number of different
words $N_d$, the total number of words $N_w$, and the success rate
$S(t)$ only depend on $t$ through the characteristic length scale
$\xi(t)$. Further, for the typical time $t_c$ to reach global
agreement or consensus, one has $\xi^d(t_c)$$\sim$$N$, i.e.,
\begin{equation}
t_c\sim N^{1/(d\gamma)} \;.
\label{tc}
\end{equation}

\subsection{Simulation Results}

\begin{figure}[t]
\vspace*{2.0truecm}
       \includegraphics{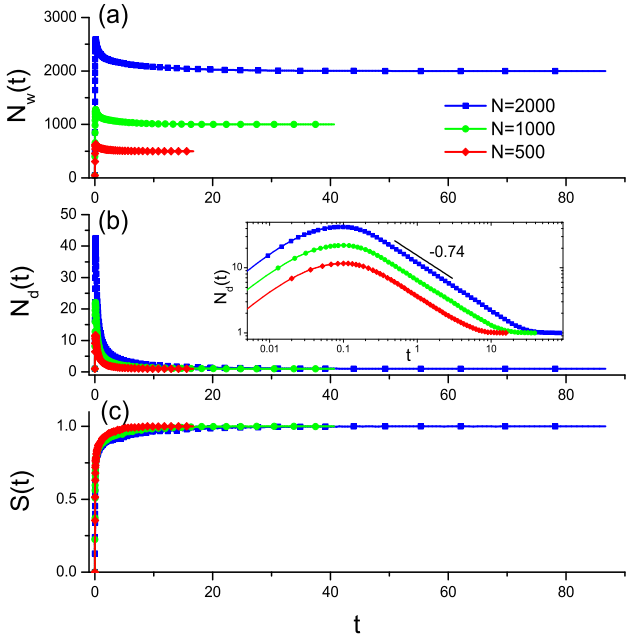}
\vspace*{7.5truecm} \caption{Time evolution of the relevant
observables in the Naming Game in  random geometric networks (RGG)
for three system sizes, averaged over $1,000$ independent network
realizations; (a) the total number of words in the
system $N_{w}(t)$; (b) the number of different words $N_{d}(t)$;
(c) the average success rate $S(t)$. The average degree of the
underlying RGGs is $\overline{k}$$\approx$$50$. The inset of (b)
shows $N_{d}(t)$  on log-log scales, displaying the late-stage
coarsening and the corresponding power-law decay, approximately
$N_d(t)$$\sim$$t^{-0.74}$.}
\label{fig.nd-nw-rgg}
\end{figure}
\begin{figure}[t]
\vspace*{2.0truecm}
       \includegraphics{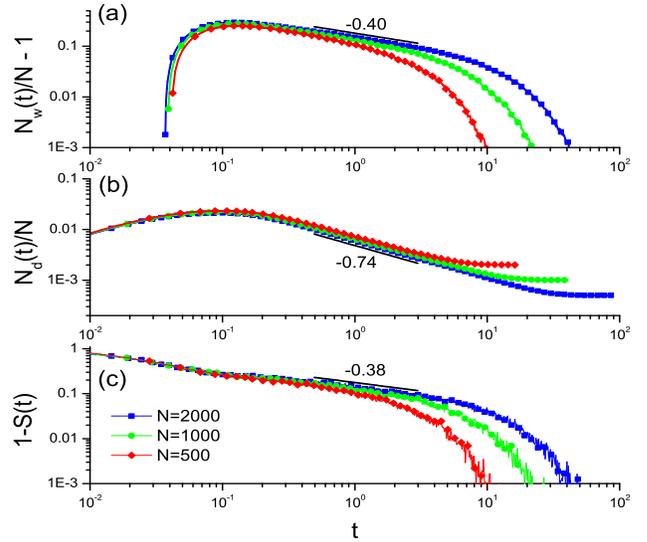}
\vspace*{5.0truecm} \caption{The scaled version of the same data
shown in Fig.~\ref{fig.nd-nw-rgg} on log-log scales; (a) the total
number of words in the system $N_{w}(t)/N$$-$$1$; (b) the number
of different words $N_{d}(t)/N$; (c) the average success rate
$1$$-$$S(t)$. The straight line segments correspond to the
best-fit power-law decays $N_{w}(t)/N$$-$$1$$\sim$$t^{-0.40}$,
$N_d(t)/N$$\sim$$t^{-0.74}$, $1$$-$$S(t)$$\sim$$t^{-0.38}$ for
(a), (b), and (c), respectively.}
\label{fig.nd-nw-rgg-scaled}
\end{figure}

Relevant quantities measured in the simulations are the total
number of words in the system $N_{w}(t)$ (corresponding to the
total memory used by the agents for word allocation at time $t$),
the number of different words $N_{d}(t)$, and the average rate of
success $S(t)$ of the word-matching attempts.
Figure~\ref{fig.mf-rgg-reg} displays the time evolution of these
three quantities for the RGG, compared to the fully connected (FC)
and to the $2d$ regular networks. Here, for the comparison, we
reproduced the corresponding data of
Refs.~\cite{Baronchelli_2005a,Baronchelli_2006a}. The behavior of
the NG on RGG is qualitatively very similar to that of the NG on
$2d$ regular graphs. After time of ${\cal O}(1)$, ${\cal
O}(N/(\overline{k}+1))$ different words have been invented
[Fig.~\ref{fig.mf-rgg-reg}(b) and \ref{fig.nd-nw-rgg}(b)].
$N_w(t)$ also reaches its maximum in time of ${\cal O}(1)$
[Fig.~\ref{fig.mf-rgg-reg}(a) and \ref{fig.nd-nw-rgg}(a)]].

Focusing on the late-time behavior of the systems, plotting
$N_d(t)/N$, $N_w(t)/N$$-$$1$, and $1$$-$$S(t)$ vs $t$ on log-log
scales, confirms the power-law decays associated with the
underlying coarsening dynamics, predicted by Eqs.~(\ref{Nd}),
(\ref{Nw}), and (\ref{S}), respectively.

\begin{figure}[t]
\vspace*{2.0truecm}
       \includegraphics{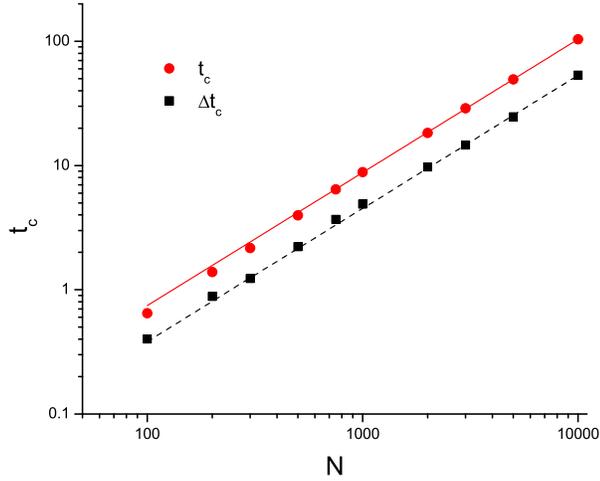}
\vspace*{4.0truecm}
\caption{Average and the standard deviation of the convergence time $t_c$ until
global agreement is reached, as a function of the number of nodes on log-log scales,
averaged over $1,000$ independent realizations of the RGG.
The average degree of the underlying RGGs is $\overline{k}$$\approx$$50$.
The straight lines correspond to the best-fit power-laws with exponent $1.07$
for both the average (solid line) and for the standard deviation (dashed line).}
\label{fig.conv-max-rgg}
\end{figure}

Further, in the scaling regime [Figs.~\ref{fig.nd-nw-rgg-scaled}(a) and (c)] we find
two consistent estimates for the exponent
of the typical length scale [see Eqs.~(\ref{Nw}) and (\ref{S})]:
$\gamma$$\approx$$0.40$ and $\gamma$$\approx$$0.38$, respectively.
The number of different words, according to Eq.~(\ref{Nd}), in
turn, should scale as $N_d(t)/N$$\sim$$1/t^{2\gamma}$, close to
our measured exponent $2\gamma$$\approx$$0.74$
[Figs.~\ref{fig.nd-nw-rgg}(b) and \ref{fig.nd-nw-rgg-scaled}(b)].
The time to global agreement scales as $t_c$$\sim$$N^{1.07}$,
shown in Fig.~\ref{fig.conv-max-rgg}, somewhat deviating from the
one predicted by Eq.~(\ref{tc}) with the exponent $1/(2\gamma)$.
This deviation is possibly due to the presence of multiple length and time scales
in the late stage of the agreement dynamics and finite-size effects.

In addition to the average convergence time $t_c$, we also
measured the standard deviation $\Delta t_c$
[Fig.~\ref{fig.conv-max-rgg}]. These results also indicate some
weakness of the NG from a system-design viewpoint: up to the
system sizes we could simulate, the standard deviation, within
error, scales in the same fashion with the number of nodes as the
average itself, $\Delta t_c$$\sim$$N^{1.07}$
[Fig.~\ref{fig.conv-max-rgg}]. The lack of self-averaging for
large systems (strong dependence on the individual runs) of the NG
is inherently related to the coarsening dynamics, having a single
interface wondering at the latest stage (and not to the underlying
random structure). Suppressing large average convergence times and
the corresponding large standard deviations will be addressed in
the next section.

\section{Naming Games in Small-World-Connected Random Geometric Networks}

In light of recent results on NG on one-dimensional SW networks
\cite{Baronchelli_2006b}, we now consider accelerating the
agreement process by adding random long-range communication links
between a small fraction of nodes of the RGG. Such networks have
long been known to speed up the spread of local information to
global scales \cite{Watts98,Watts99,NEWMAN_SIAM,KHK_PRL2005}, with
applications ranging from synchronization problems in distributed
computing \cite{KNGTR03a} to alarm-detection schemes in wireless
sensor networks \cite{LKS_INSS2006}. For sensor networks, this can
be implemented either by {\em adding} a small fraction of sensors
equipped with long-range unidirectional antennas (``physical"
long-range connections) or by establishing designated multi-hop
transmission patterns (``logical" long-range connections) between
certain nodes \cite{HELMY_2003}.

We construct the small-world-like RGG (SW RGG) as follows. We
start with the original RGG (embedded in $d$ dimensions, where
$d$$=$$2$ in this paper). Then we {\em add} ``long-range" links (or ``shortcuts")
between randomly chosen nodes in such a way that the
total number of long-range links per node (the density of random
links) is $p$.  This SW construction slightly differs from the
original Watts-Strogatz one \cite{Watts98} [also used by Dall'Asta
et al. \shortcite{Baronchelli_2006b}], where random links are
introduced by ``rewiring" some of the original connections. The
resulting network, however, has the same universal properties in
the small-$p$, large-$N$ limit \cite{NW_1999}, which is the center
of our interest. Further, it is also motivated by actual
implementations in sensor networks.

\subsection{Basic Scaling Considerations}

Before presenting simulation results, using scaling arguments, one
can obtain an order of magnitude estimate for the crossover time
$t_{\times}$ present in the SW RGG and for the time to reach
global agreement $t_c$ \cite{Baronchelli_2006b}. In SW networks,
embedded in $d$ dimensions, the typical distance between two nodes
with an added long-range link emanating from them scales as
$l_{SW}$$\sim$$p^{-1/d}$. Starting from empty initial word lists,
for early times (following the creation of ${\cal
O}(N/(\overline{k}+1))$) different words in the system), the
system will exhibit  coarsening, until the typical size of the
growing domains $\xi(t)$$\sim$$t^{\gamma}$ becomes comparable to
$l_{SW}$. After that time, the agreement process is governed by
the presence of random long-range connections, yielding
mean-field-like behavior. Hence the crossover from $d$-dimensional
coarsening to mean-field-like dynamics occurs when
$t^{\gamma}$$\sim$$p^{-1/d}$, yielding
\begin{equation}
t_\times\sim p^{-1/(d\gamma)}\;. \label{t_x}
\end{equation}
In a system of $N$ agents, the above crossover is only displayed
if the convergence time of the original system with no random
links would exceed the above crossover time
$N^{1/d\gamma}$$\gg$$p^{-1/d\gamma}$, which is equivalent to the
condition for the onset of the SW effect $N$$\gg$$p^{-1}$
\cite{NW_1999,Baronchelli_2006b}. Following the above system-size
independent crossover time, the agreement dynamics is of
mean-field like, and one can expect to observe a scaling behavior
close to that of FC networks \cite{Baronchelli_2005a}. In
particular, the time to reach global agreement is expected to
scale as
\begin{equation}
t_c\sim N^{1/2}
\label{t_cmf}\;,
\end{equation}
a significant [and anticipated \cite{Baronchelli_2006b}] reduction
compared to that of the ``pure" RGG with no long-range links where
$t_c$$\sim$$N^{1.07}$.

\subsection{Simulation Results}

\begin{figure}[t]
\vspace*{2.0truecm}
       \includegraphics{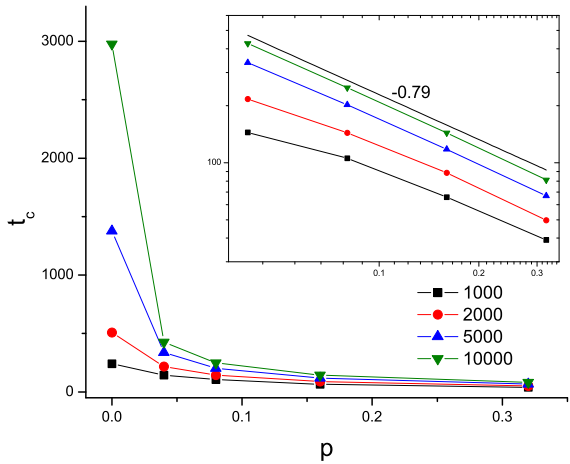}
\vspace*{4.0truecm}
\caption{Average  convergence time $t_c$ for
SW RGGs, as a function of the density of shortcuts for various
system sizes, averaged over $1,000$ independent realizations of
the network. The average degree of the underlying RGGs is
$\overline{k}$$\approx$$12$. The inset shows the same data on
log-log scales. The straight lines corresponds to an estimate of
the associated (asymptotic) power-law.}
\label{fig.conv-p-swrgg}
\end{figure}

Simulating the NG on SW RGGs confirms the above scaling scenario.
Following the very early-time development of ${\cal
O}(N/(\overline{k}+1))$ different words, the system of
SW-networked agents, exhibits slow coarsening, with only small
corrections to the behavior of the pure RGG. In fact, this
early-time coarsening on SW RGGs is slightly slower compared to
pure RGGs due to the effective pinning of interfaces near the
shortcuts
\cite{Baronchelli_2006b,Boyer2003,Castellano2003,Castellano2005}.
In the NG on SW networks, however, the agreement process only
slows down \cite{Baronchelli_2006b}, but is not halted by
``frozen" (metastable) disordered configurations
\cite{Boyer2003,Castellano2005}. After a $p$-dependent crossover
time [Eq.~(\ref{t_x})], (when the typical size of the growing
clusters becomes comparable to the SW length scale), an
exponential convergence begins to govern the agreement process.
This final-stage fast approach toward consensus sets in earlier
for increasing values of the density of shortcuts $p$, yielding a
significantly reduced convergence time compared to that of the NG
on the ``pure" RGG.
Plotting the convergence time vs the density of long-range links,
as shown in Fig.~\ref{fig.conv-p-swrgg}, suggests that (for
sufficiently large but {\em fixed} $N$) the convergence time
approaches an asymptotic power-law $t_{c}$$\sim$$p^{-s}$ with
$s$$\approx$$0.79$.

\begin{figure}[t]
\vspace*{2.0truecm}
       \includegraphics{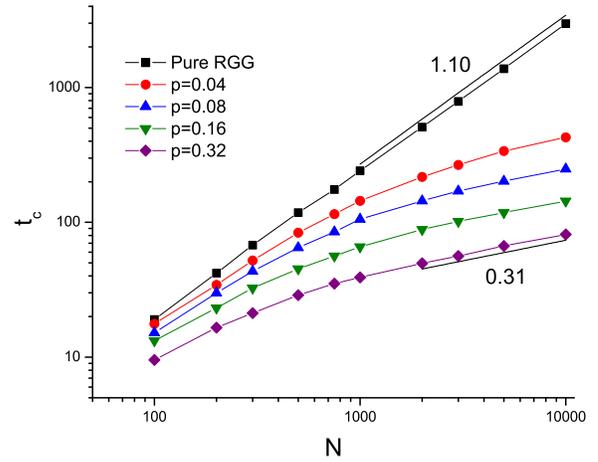}
\vspace*{4.0truecm}
\caption{Average  convergence time $t_c$
for SW RGGs, as a function of the number of nodes on log-log
scales for various density of long-range links $p$, averaged over
$1,000$ independent realizations of network. The average degree of
the underlying RGGs is $\overline{k}$$\approx$$12$. The straight
line segments correspond to the best-fit (asymptotic) power-laws
with exponents $1.10$ and $0.31$, for the ``pure" RGG ($p$$=$$0$) and for the SW RGG
($p$$>$$0$) cases, respectively.}
\label{fig.conv-N-swrgg}
\end{figure}

For fixed $p$ and increasing $N$, however, the convergence time
still increases with $N$ [Fig.~\ref{fig.conv-N-swrgg}],
$t_c$$\sim$$N^{0.31}$,
closer to the anticipated mean-field-like behavior [Eq.~(\ref{t_cmf})]
\cite{Baronchelli_2006b}.

\section{Summary and Outlook}

In this paper, we have explored the Naming Games on Random
Geometric Graphs and SW-connected RGGs. While the underlying RGG
communication topology is motivated by large-scale sensor
networks, the NG on (SW) RGGs captures fundamental features of
agreement dynamics of spatially-embedded networked agent-based
systems. We have found that, qualitatively similar to
two-dimensional regular networks \cite{Baronchelli_2006a}, the NG
on RGG can be reasonably well described by the physical theory of
coarsening. In particular, local clusters of nodes sharing the
same word quickly form, followed by slow coarsening of these
clusters in the late stage of the dynamics.
Our simulation results indicate that the
average time to reach global agreement is of ${\cal O}(N^{1.07})$
(for {\em fixed} average degree). By adding a small density of
shortcuts on ``top" of the RGG, resulting in a SW-like network,
the convergence time is strongly reduced and becomes of ${\cal
O}(N^{0.31})$, closer to that of the FC network \cite{Baronchelli_2006b}.

In future work we will investigate the NG on more realistic
communication topologies, motivated by and relevant to wireless
sensor networks, in particular, random spatial networks with
heterogeneous range distribution, and also networks with
dynamically changing connectivity.


\end{document}